\documentclass[a4paper,conference]{IEEEtran}
\IEEEoverridecommandlockouts
\usepackage{cite}
\usepackage{amsmath,amssymb,amsfonts}
\usepackage{algorithmic}
\usepackage{graphicx}
\usepackage{textcomp}
\usepackage{xcolor}
\usepackage{orcidlink}
\usepackage{kotex}
\usepackage{lipsum}
\usepackage{booktabs}
\usepackage{graphicx}
\usepackage{romannum}
\usepackage{amsmath}
\usepackage{kotex} 
\usepackage[algo2e, ruled, vlined, linesnumbered]{algorithm2e}
\usepackage{algorithm}
\usepackage{xcolor}
\usepackage{graphicx}

\def\BibTeX{{\rm B\kern-.05em{\sc i\kern-.025em b}\kern-.08em
    T\kern-.1667em\lower.7ex\hbox{E}\kern-.125emX}}

\begin{document}

\newcommand{\blu}{\color{blue}}
\newcommand{\blk}{\color{black}}
\newcommand{\qes}{\color{red}}

\title{Generative Active Learning with Variational Autoencoder for Radiology Data Generation\\ in Veterinary Medicine \\

}

\author{\IEEEauthorblockN{In-Gyu Lee} \IEEEauthorblockA{\textit{Dept. Computer Science} \\ \textit{Chungbuk National University}\\ 
Cheongju, Republic of Korea  \\ 
ingyu.lee@chungbuk.ac.kr} 
\\ 
\IEEEauthorblockN{Jae-Hwan Kim} \IEEEauthorblockA{\textit{Dept. Veterinary Medical Imaging} \\ \textit{Konkuk University}\\ 
Seoul, Republic of Korea \\ 
jaehwan@konkuk.ac.kr} 
\and 
\IEEEauthorblockN{Jun-Young Oh} \IEEEauthorblockA{\textit{Dept. Computer Science} \\ \textit{Chungbuk National University}\\ 
Cheongju, Republic of Korea \\ 
jy.oh@chungbuk.ac.kr} 
\\ 
\IEEEauthorblockN{Ki-Dong Eom} \IEEEauthorblockA{\textit{Dept. Veterinary Medical Imaging} \\ \textit{Konkuk University}\\ 
Seoul, Republic of Korea \\ 
eomkd@konkuk.ac.kr} 
\and 
\IEEEauthorblockN{Hee-Jung Yu} \IEEEauthorblockA{\textit{Dept. Veterinary Medical Imaging} \\ \textit{Konkuk University}\\ 
Seoul, Republic of Korea \\ 
hazel13@konkuk.ac.kr} \\ 

\IEEEauthorblockN{Ji-Hoon Jeong*} \IEEEauthorblockA{\textit{Dept. Computer Science} \\ \textit{Chungbuk National University}\\ 
Cheongju, Republic of Korea \\ 
jh.jeong@chungbuk.ac.kr}}
\maketitle

\begin{abstract}
Recently, with increasing interest in pet healthcare, the demand for computer-aided diagnosis (CAD) systems in veterinary medicine has increased. The development of veterinary CAD has stagnated due to a lack of sufficient radiology data. To overcome the challenge, we propose a generative active learning framework based on a variational autoencoder. This approach aims to alleviate the scarcity of reliable data for CAD systems in veterinary medicine. This study utilizes datasets comprising cardiomegaly radiograph data. After removing annotations and standardizing images, we employed a framework for data augmentation, which consists of a data generation phase and a query phase for filtering the generated data. The experimental results revealed that as the data generated through this framework was added to the training data of the generative model, the frechet inception distance consistently decreased from 84.14 to 50.75 on the radiograph. Subsequently, when the generated data were incorporated into the training of the classification model, the false positive of the confusion matrix also improved from 0.16 to 0.66 on the radiograph. The proposed framework has the potential to address the challenges of data scarcity in medical CAD, contributing to its advancement.
\end{abstract}


\section{Introduction} 
Pets have become important members of our lives, forming strong bonds and emotional connections with their owners. The healthcare of pets has become a subject of increased interest. With the continuous evolution of artificial intelligence (AI), there is a noticeable trend towards incorporating AI into computer-aided diagnosis (CAD) systems for pet healthcare. The effectiveness of AI models is heavily dependent on access to high-quality training data \cite{zhu2016we,gudivada2017data}. However, acquiring a substantial amount of medical data for CAD has challenges due to the sensitive personal information in such data. Consequently, there is a persistent effort to explore the application of generative models for the creation of medical data.

Among these efforts, numerous studies have leveraged generative adversarial networks (GAN) \cite{Oh2023Application}. Yoon et al. \cite{yoon2022colonoscopic} achieved a fretchet inception distance (FID) of 42.19 for Sessile serrated lesion images using style-based GAN. Salvia et al. \cite{la2022deep} proposed the use of GAN to generate synthetic hyperspectral images of epidermal lesions, addressing the challenge of limited large datasets. These academic efforts demonstrate the potential of GAN in generating medical images to improve research and diagnostics.

Zhu et al. \cite{zhu2017generative} explored the concept of generative adversarial active learning (GAAL), employing a generative model in active learning to improve the performance of classification models. Although this approach involved queries to augment the training dataset for labeling, they emphasized the proposed framework, rather than focusing on criteria for queries.

In this study, we have used creatively used active learning the context of generative models, incorporating a unique criterion for data filtration through a variational autoencoder (VAE) \cite{kingma2013auto}. This approach has significantly enhanced the robustness of the generative model's performance. We focused on addressing the scarcity of medical data for CAD, especially in the field of veterinary medicine, which may be helpful in medical AI advances. The introduced approach, termed Generative Active Learning with VAE, leverages query processes facilitated by a VAE to improve the performance of generative models in generating medical image data. This method stands out as a viable solution to address the persistent challenge of limited medical image data in CAD applications.

\begin{figure}[!t]
\centering
\includegraphics[width=\columnwidth]{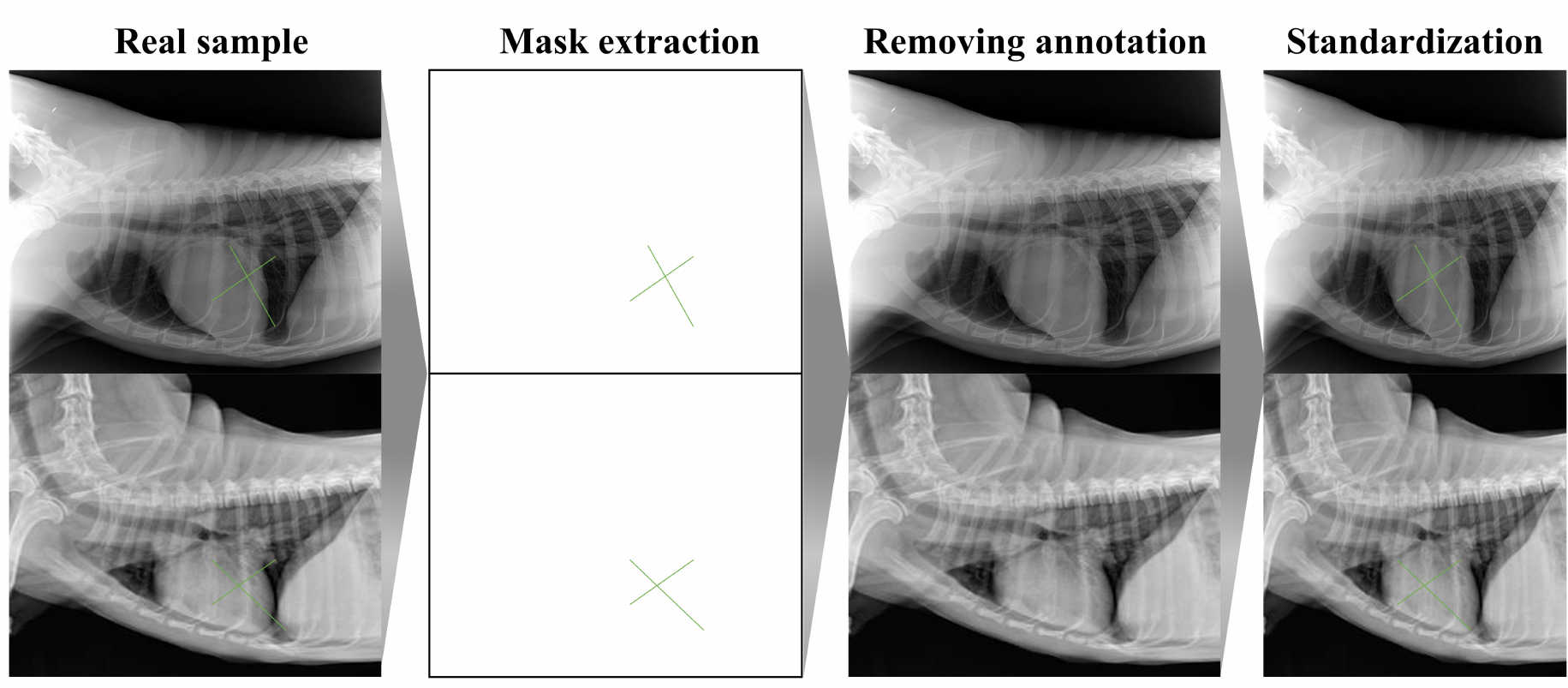}
\caption{The data preprocessing pipeline for training a generative model. If doctors draw annotations for diagnosis, the annotations are extracted to create masks. Then, image inpainting techniques are applied to remove the annotations. Subsequently, the resolution of the images is standardized.}
\end{figure}

\section{Generative Active Learning with VAE}
\subsection{Dataset}
In this study, we utilized the ``Image Data for Diagnosis of Pet Diseases (thorax)'' from AIHUB, a public dataset. We selected 100 images from cardiomegaly disease data. Data on cardiomegaly disease can visually confirm that the heart is enlarged \cite{lam2021radiographic}. We used these selected data as initial training data for the generative model.

Before training the model, a preprocessing step was performed to improve the quality of the data used for training. For radiographic images, annotations made by veterinarians for diagnosis were present. As the model could potentially learn from these annotations as features of the data, we performed a task to remove the annotations. Initially, the data in RGB format was transformed into the HSV color space. Subsequently, since the color of the annotations was in kinds of green, we extracted the green tones to create a binary mask. The mask obtained was then utilized in the image inpainting technique to restore the image. This method replaces pixels in the masked area using neighboring pixels.

Second, we standardized the resolution of both radiographic images. The raw data had diverse resolutions. Inconsistent image resolutions in the training dataset can lead to unstable learning due to variations in the size of the feature maps extracted by the neural network. To address this, we employ the center-cropping method, which uses center-based image cropping to ensure that essential organ information, such as the heart and kidney, is not lost. Radiographic images were resized to 256$\times$256 pixels. The data preprocessing procedure is depicted in Fig. 1.

\let\oldnl\nl 
\newcommand{\nonl}{\renewcommand{\nl}{\let\nl\oldnl}}
\SetKwInput{KwInput}{Input}
\SetKwInput{KwOutput}{Output}
\SetKwInput{KwStep}{Step 1} \SetKwInput{KwStepp}{Step 2} \SetKwInput{KwSteppp}{Step 3} \SetKwInput{KwStepppp}{Step 4}

\begin{algorithm}[!t]
\linespread{1.3}
\footnotesize
\caption{Overall process of proposed framework}

\KwStep{Data Preprocessing}
\KwInput{$Raw \_ dataset$}
\KwOutput{$Preprocessed \_ dataset$}
        \If{Annotations exist in $Raw \_ dataset$}{\;
            Convert RGB images to HSV images \\
            Extract green tones to create masks \\
            $Inpainted \_ dataset =$ Use masks for image inpainting\;
        }
        $Preprocessed \_ dataset =$  Standardize resolution of $Inpainted \_ dataset$\; \\

\KwStepp{Data Generation}
\KwInput{$Preprocessed \_ dataset$} 
\KwOutput{$Augmented \_ dataset$}
        \While{$Augmented\_dataset \  size < 500$}{
            \For{$epochs = \  1$  to $20$} {
                \If{$epochs$ is even}{\;
                    $New\_FID =$ Evaluate the generative model\; \\
                    \If{$Saved\_FID > New\_FID$ }{\;
                        $Saved\_FID = New\_FID$ \\
                        $Saved\_weights =$ Save the generative model's weights
                    }
                }
            }
            $Generated\_dataset =$ Generate 1000 data using $Saved\_weights$\;
            
            Calculate the cosine similarity of $Generated\_dataset$\;
            
            $Filtered\_dataset$ $=$ Select the top 10\% $Generated\_dataset$
            
            $Augmented\_dataset += Filtered\_dataset$ \;
        } 
\end{algorithm}

\begin{figure*}[!t]
\centering
\includegraphics[width=\textwidth]{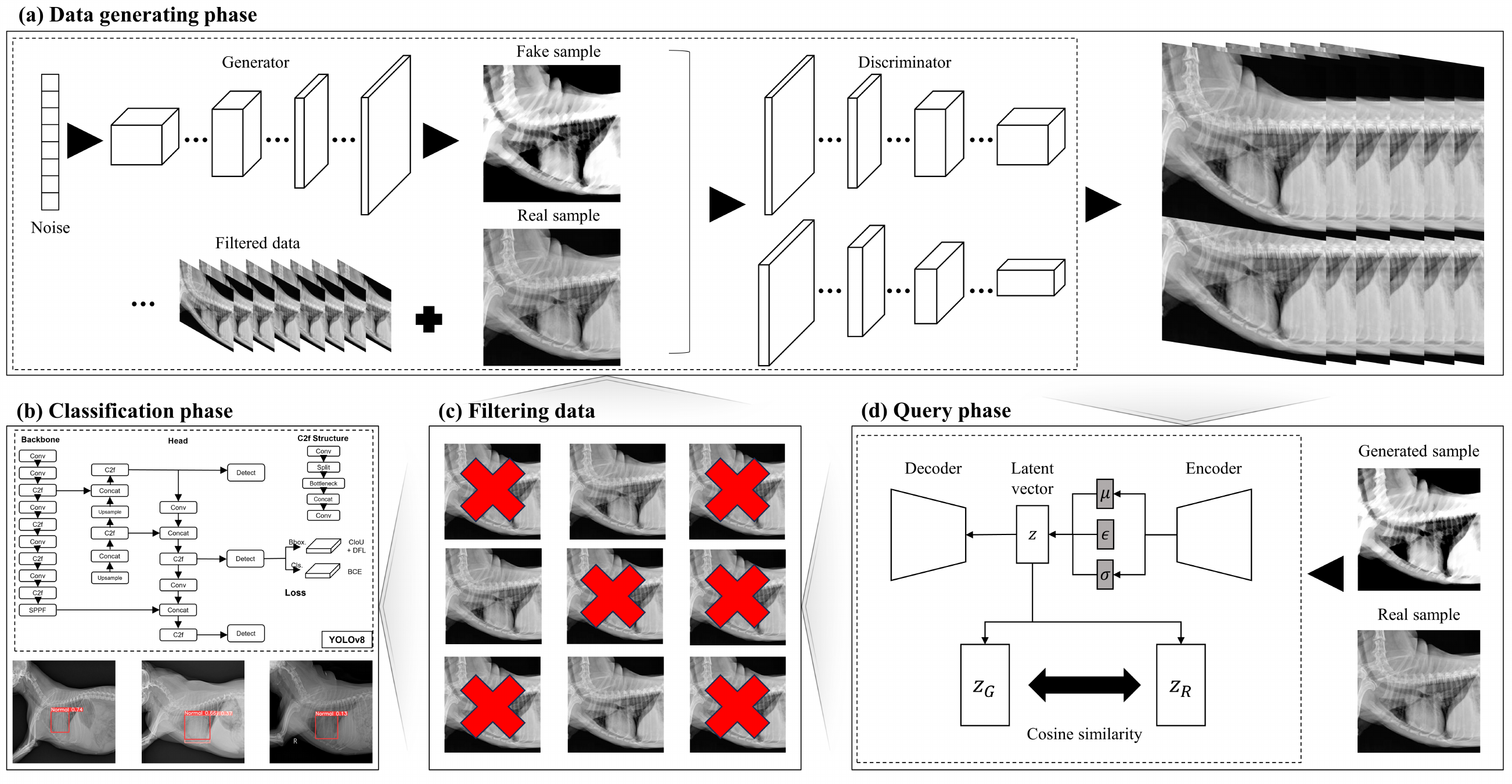}
\caption{Overall flow of the proposed framework. (a) The projectedGAN is trained with filtered image data and real image data to generate a new radiographic image. (d) VAE trained with 100 original data are used to filter generated images using a query strategy. (c) The top 10\% cosine similarity of the data is added to the training dataset. (b) Finally, classification is performed using the object detection model after labeling to prove the usefulness of the data.}
\end{figure*}

\subsection{Proposed Framework}
The framework is composed of the two phases. First, the data generating phase trains the generative model and generates data. Second, the query phase filters the generated data through the query strategy before incorporating them into the dataset for training the generative model. Algorithm 1 details the specific steps involved in this process with data preprocessing.

\subsubsection{Data generating phase}
The overall flow is depicted in Fig. 2. We refer to this entire process as a cycle and repeat the cycle 4 times until the train dataset has 500 data. The study utilized the projectedGAN model proposed by Sauel et al. \cite{sauer2021projected} due to its state-of-the-art performance across various datasets during the experimentation. ProjectedGAN comprises a generator and a discriminator. The generator is trained to learn the data features to generate images that can effectively deceive the discriminator. The discriminator learns the data in a way that distinguishes between real data and generated data. The evaluation of the generation model was based on the FID \cite{heusel2017gans}, which measures the dissimilarity of the characteristics between the generated and actual images. A lower FID value indicates superior performance. The formula to calculate the FID is provided below. $T$ represents actual images, and $G$ represents generated images. $Tr$ is defined as the sum of elements from the upper left to the lower right of the vector.

\begin{align} 
FID = \left\|\mu_{T} - \mu_{G}\right\|^{2} - Tr((\Sigma_{T} + \Sigma_{G} - 2(\Sigma_{T}\Sigma_{G})^{\frac{1}{2}})
\end{align}

The GAN initiates training by using 100 selected actual data in the initial cycle. Training progresses through a total of 20 epochs per cycle, with a performance evaluation conducted every 2 epochs. Consequently, each cycle yields a total of 10 FID assessments. During evaluation, if the current FID value is lower than the previously recorded FID value, we save the model's weights. We utilize these saved weights to generate 1,000 images per cycle.

\subsubsection{Query phase}
The evaluation of image similarity in this study used a VAE, comprising an encoder and a decoder. In VAE, the decoder aims to regenerate the input in a form that is most similar to when a latent vector is given. The encoder, on the other hand, seeks to find the mean and standard deviation of the input and generates a latent vector with noise epsilon in Gaussian distribution. The study focused on utilizing the latent vector generated by the encoder. Training the VAE involved using 100 selected actual data for a total of 25 epochs. Training the VAE involved using 100 selected actual data for a total of 25 epochs.

To assess image similarity, cosine similarity was employed \cite{rahutomo2012semantic}. Unlike distance measures such as the Euclidean distance, which evaluate vectors based on their magnitudes, cosine similarity examines whether both vectors are aligned in the same direction. This characteristic makes cosine similarity particularly suitable for gauging significant similarities between images. The formula for cosine similarity is provided below. In the given equations, \textit{T} denotes the latent vector of the true image, whereas \textit{G} represents the latent vector of the generated image.

\begin{align} 
 Cosine\ similarity=1-\frac{T\times G}{\left\| T \right\|\left\| G \right\|}
\end{align} 

The original images and the images generated through the data generating phase were passed to the autoencoder's encoder to obtain embeddings. The cosine similarity between the 100 original images and the generated image was calculated. The generated images with the top 10\% of cosine similarity were selected and added to the training set.

\subsection{Classification phase}
To demonstrate the validity of our framework, we applied a classification model to images generated using our framework. The model we used for this purpose is YOLOv8, which is an enhancement of YOLOv5 based on additional layer modifications to improve the model's performance, achieving state-of-the-art results. The YOLO series is a well-known model extensively utilized in various CAD applications \cite{10195094,10195057}.

The training dataset comprises a total of 10 sessions, divided into 5 sessions for the heart and 5 sessions for the kidney. We initiated the training with 100 samples and gradually increased the training dataset size to 500 samples. Since all data used to train the generation model are disease-related, we also labeled normal data to assess classification accuracy. Each class, including normal, was labeled with 500 samples per class. For testing, we extracted 50 data samples per class from the actual data that were not duplicated with the training data.

\begin{table*}[t]
\centering
\caption{The results of FID values for generating radiographic image}
\resizebox{\linewidth}{!}{%
\begin{tabular}{cccccccccc}
\toprule
\multicolumn{2}{c}{\textbf{Original}} & \multicolumn{2}{c}{\textbf{Cycle-1}} & \multicolumn{2}{c}{\textbf{Cycle-2}} & \multicolumn{2}{c}{\textbf{Cycle-3}} & \multicolumn{2}{c}{\textbf{Cycle-4}} \\
\textbf{Optimal}   & \textbf{Worst}   & \textbf{Optimal}   & \textbf{Worst}  & \textbf{Optimal}   & \textbf{Worst}  & \textbf{Optimal}   & \textbf{Worst}  & \textbf{Optimal}   & \textbf{Worst}  \\ \midrule
84.14              & 100.16           & 64.31              & 97.21           & 58.39              & 82.23           & \textbf{50.75}              & \textbf{74.57}           & 53.87              & 81.82           \\ \bottomrule
\end{tabular}%
}
\end{table*}

\begin{figure*}[!t]
\centering
\includegraphics[width=\linewidth]{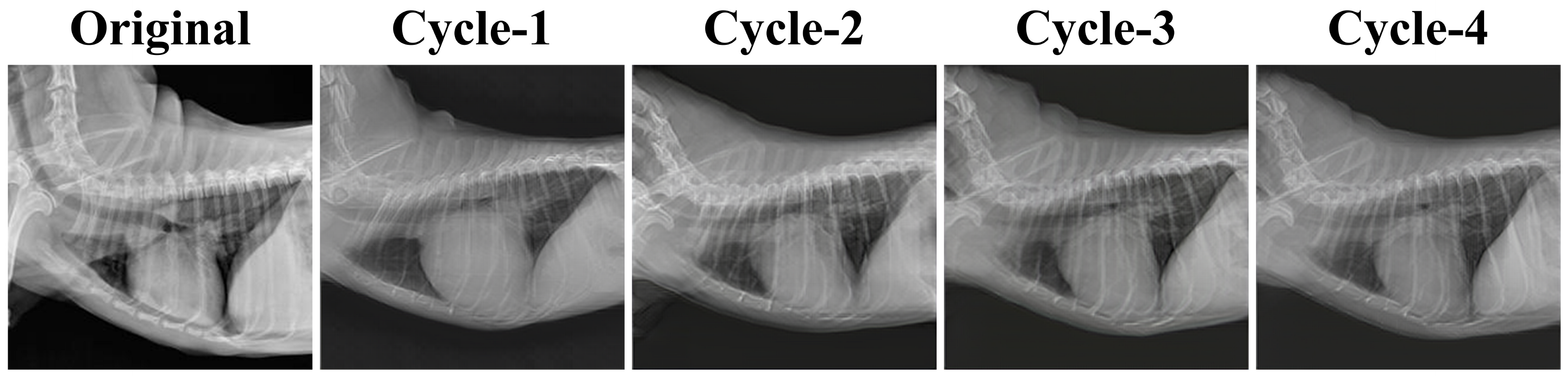}
\caption{Generation results of each cycle. The following is data generated by learning cardiomegaly data.}
\end{figure*}

For evaluation metrics, we utilized the confusion matrix, along with accuracy, precision, recall, and F1-score \cite{powers2020evaluation,tharwat2020classification}. The confusion matrix is a table used in machine learning to assess the performance of a classification model, summarizing the relationship between the model's predictions and actual values. From this matrix, accuracy, precision, recall, and F1-score can be calculated using the following formulas. In these formulas, true positive (TP) represents the number of correctly predicted positive observations, while true negative (TN) denotes the number of correctly predicted negative observations. False positive (FP) indicates instances predicted as positive, but actually negative. False negative (FN) indicates the number of instances that are actually positive but are incorrect. 

\begin{align} 
Accuracy: \frac{TP + TN}{TP + TN + FP + FN}
\end{align} 

\begin{align} 
Precision: \frac{TP}{TP + FP}
\end{align} 

\begin{align} 
Recall: \frac{TP}{TP + FN}
\end{align} 

\begin{align} 
F1 \ score: \frac{2 \cdot (Precision \cdot Recall)}{Precision + Recall}
\end{align}

\begin{figure*}[!t]
\centering
\includegraphics[width=\textwidth]{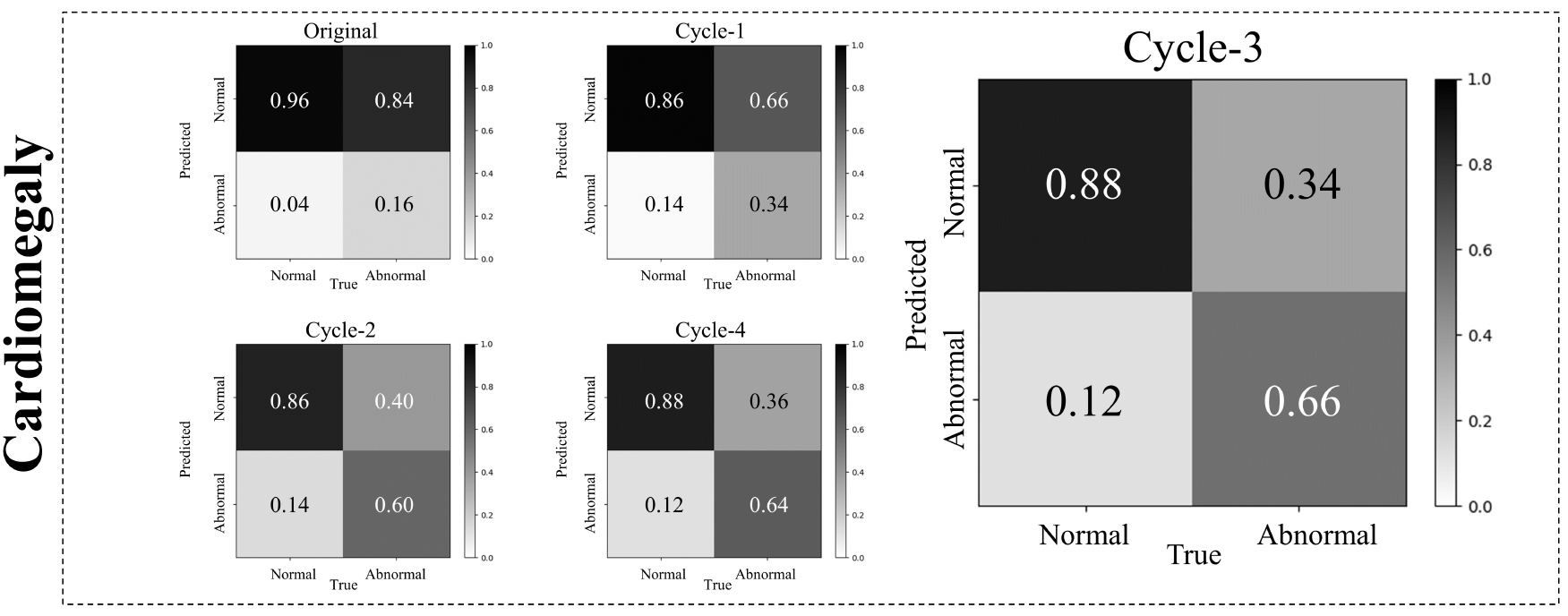}
\caption{Confusion matrix results of classification phase. The above results represent training and testing results using cardiomegaly data. Among the five sessions, the confusion matrix of the session with the highest accuracy is presented more prominently.}
\end{figure*}



\section{Experiments}
\subsection{Data generating phase}

The results of the FID experiment to generate radiograph data are presented in Table I. In total, there are five sessions, ranging from original to cycle-4, each divided into `Optimal' and `Worst' cases. The original session involves training the model exclusively on selected data from the original dataset. `Optimal' represents the lowest FID value among the 10 values, while `Worst' indicates the highest FID value out of the 10. A lower FID value implies better performance of the generative model.

The term cycle refers to the process of generating data in the data generating phase, filtering through the query phase, and adding the filtered data to the training dataset of the generative model. The number following the cycle represents the iteration count. For example, cycle-1 involves training the generative model with 100 new data added to the training dataset, generated using the model trained on the data from the original session. Consequently, in cycle-1, the size of the training dataset is 200. Subsequently, cycle-2 includes the 200 data previously used and an additional 100 generated data.

In cycle-3, the generated radiograph data showed the best performance with the `Optimal' score of 50.75 and the `Worst' performance at 74.57. Additionally, a trend was observed in which performance tended to be less favorable with a smaller amount of data. On the contrary, as the amount of data increased, there was a performance improvement, although the final cycle did not consistently yield the best results for radiographic images. The generated data examples are shown in Fig. 3, where the left side displays the original data and the results obtained through cycle-4. In Fig. 3, the displayed results represent the filtered data for each cycle with the highest cosine similarity.

\subsection{Classification phase}
The results of the classification using YOLOv8 are presented in Fig. 4. Fig. 4 includes the confusion matrix, where the upper part represents the results tested on radiographs of dogs with cardiomegaly. The confusion matrix is commonly employed as an evaluation metric in various research studies involving classification tasks \cite{9920692}. Each classification task was conducted to demonstrate the validity of the data and the classification model training utilized the dataset used for the generative model.

In cardiomegaly data, cycle-3 showed the highest accuracy and precision, with values of 0.73 and 0.72, respectively. In addition, cycle-3 and cycle-4 showed the highest F1-score of 0.79. The recall value was highest when trained with the original data.

On the other hand, in the case of cardiomegaly, the session with the lowest accuracy was not the original, but cycle-1. In cycle-1, the accuracy was 0.66, which was 0.01 lower than the accuracy of 0.67 in the original session.

\section{Discussion}
In this paper, we propose a generative active learning framework that automates the query process using the VAE. This framework generates data during the data generating phase and incrementally augments the training dataset of the generative model by filtering data through the query phase. Unlike previous research, we adopt the VAE to enhance the robustness of the query process. The query process has the filtering step by calculating the cosine similarity between the generated images and real images using 10\% of the generated data. Experimental results demonstrate that iterative repetition of this process leads to improved performance of the generative model. 

Observing the change in FID during the data generating phase, there was a consistent trend of FID reduction as the cycles progressed. For each session, 10 FID scores were obtained. In the case of radiographic data generation, the `Optimal' FID score decreased from 84.14 to 50.75, and the `Worst' FID score decreased from 100.16 to 74.57 throughout the cycles. These results demonstrate that our proposed framework effectively enhances the robustness of the generative model's performance.

To validate the validity of our data, we conducted a classification phase. Data used for generation belonged to all categories of disease, including cardiomegaly. The experimental results, as observed from the confusion matrix, reveal that as cycles increase, the increase in disease data leads to an increase in FP. In the case of cardiomegaly, FP increased from 0.16 to 0.66. However, the overall accuracy for both diseases reached its highest value at cycle-3. It is important to note that when evaluating the model's performance numerically, the classification performance of normal data also influences the results. Therefore, while accuracy has increased slightly, the significant increase in FP suggests that the generated data using our framework has demonstrated its utility in improving the performance of the classification model.

Furthermore, this study has some limitations. First, our methodology involved the utilization of an existing GAN variant instead of the proposed model. In particular, contemporary image generation models lean toward diffusion models \cite{ho2020denoising,song2020denoising} rather than GAN. Second, while there is a plethora of medical image data available, we restricted our application of the framework to radiographic image data. Given the limited dataset that encompasses only these two modalities, further exploration is essential across diverse datasets to establish the generalizability of our findings. These limitations highlight avenues for future research and improvements in our approach.

\section{Conclusion and Future works}
This study proposed the VAE-based generative active learning framework. The potential of this framework to address the issue of medical data scarcity in CAD was demonstrated through experimental results, including the FID of the generative model and the confusion matrix, accuracy, F1-score, precision, and recall of the classification model. Future research will extend to proposing generative model such as diffusion models and using various types of data, such as computerized tomography (CT), magnetic resonance imaging (MRI), etc. It will have a positive impact on the performance improvement of the CAD system in the future and provide an opportunity to promote the development of the medical AI field.

\bibliographystyle{IEEEtran}
\bibliography{ref}

\end{document}